# Constraint-based Query Distribution Framework for an Integrated Global Schema

Ahmad Kamran Malik[1], Muhammad Abdul Qadir[1], Nadeem Iftikhar[2], and Muhammad Usman[3]
[1]Muhammad Ali Jinnah University, Islamabad, Pakistan
kamranocp@gmail.com, aqadir@jinnah.edu.pk
[2]Aalborg University, Aalborg, Denmark; nadeem@cs.aau.dk
[3]New York University, New York, USA; mut201@nyu.edu

*Abstract* – **Distributed heterogeneous data sources need to be queried uniformly using global schema. Query on global schema is reformulated so that it can be executed on local data sources. Constraints in global schema and mappings are used for source selection, query optimization, and querying partitioned and replicated data sources. The provided system is all XML-based which poses query in XML form, transforms, and integrates local results in an XML document. Contributions include the use of constraints in our existing global schema which help in source selection and query optimization, and a global query distribution framework for querying distributed heterogeneous data sources.**

## I. INTRODUCTION

Large number of heterogeneous data sources are available online about every field of life. Data sources contain data in variety of formats. For example, different departments in an organization can store data in different formats like XML file, object based, or as a database. This heterogeneous distributed data need to be queried to fulfill the needs of organization. An integrated global schema of sources is required that can be queried for accessing all the data sources without knowing details of the individual sources. We use the integrated schema of sources created in our previous system [1]. Different types of query processing systems have been proposed in literature. Our system like most of the other systems [2], [3] poses query on already integrated global schema. Some other systems like [4], [5] integrate data at runtime when a query is posed without creating the global schema. The approach which does not use a global schema is suitable for sources which are changing rapidly [4]. An interesting survey of data integration systems is presented in [6].

Query posed on global schema needs to be reformulated according to the local source schema and local source format. For this purpose source selection needs to be performed which determines sources that can answer the query. Also the query needs to be evaluated on sources based on the source capability. This requires query optimization so that only that part of the query need to be executed on local source which satisfies local schema constraints. Sources can be replicated and partitioned according to the requirements of organizations. Local schema constraints, such as integrity constraints and validation rules, are matched with query constraints for source selection and query optimizations. These constraints may include length, format, and validation rules. We added these constraints in the global schema and mapping documents created by [1]. In addition to constraints, element names can differ in local and global schema. This requires mapping between global and local schema using the domain ontology. For global query to local source mappings our system uses global ontology of domain created by [1]. A survey of ontology based query systems is presented in [7].

Following are the main features of our query distribution framework. Our query distribution framework is all XML-based, as XML is becoming the standard for the data exchange. We use XPath as query language. Constraints are used in the global schema and mappings for helping in query optimization and source selection. Global ontology of sources is already created [1] for specific domain which is used for mapping between global query and sources semantics. The system is implemented using java-based Web services. Our system supports XML, Relational and Object-oriented sources. Results from heterogeneous data sources are merged in a single XML document.

Remainder of the paper is organized as follows. Section II describes related work. Section III describes constraint based global schema and mappings. Section IV explains our proposed query distribution framework. Section V presents a case study. Section VI describes the conclusion.

## II. RELATED WORK

For querying distributed heterogeneous resources, some systems use global schema which is also named as world view in [8], others use multiple independent schemas. Following systems describe query processing and reformulation techniques used in literature.

A GUI-based query interface is used by [9] which converts query into SQL language. Ontology is used to map query terms with source schema. Local queries extract individual results which are converted into XML format and integrated after eliminating conflicts using ontology. This system uses SQL language for all the query conversions while we use XML-based queries which is a standard language for data exchange.

Agora is an XML-based system which uses XQuery language to pose queries on global schema [10]. It translates

XQuery into SQL format. Some of the XQuery constructs have no equivalent in SQL, so few queries may not be translated and executed. Using LAV mappings of sources, query is rewritten according to the source schema. Agora converts XQuery into SQL while we are using XQuery and then converting it into three types of local sources.

Infomix is a system that uses constraints to check for the decidability of query answering when data is incomplete or inconsistent [2], while our system is not created for this purpose. Mapping are in GAV (Global As View) form. Global schema integrity constraints are used to reformulate the query into multiple queries destined for sources.

PIM (Personal Information Management) systems are used to manage different types of personal information as mentioned in SemanticLIFE project [11]. Queries can be posed by showing the objects of the sources using ontology. It can create more possible subqueries from the original query using ontology.

IBHIS system is a service-oriented system [3]. Query is posed on federated schema, and is decomposed into local queries which use web services to access each type of registered source. This system is created for health care domain.

XML data is described as relational views for efficiency and optimization reasons [12]. XML sources are described as views on generic relational schema. Query is cost optimized especially for data transfer cost and sent to sources, while our system uses integrity constraints and validation rules for source selection and optimization.

Unlike above described systems [9], [10] which convert XQuery into SQL and then translate and forward query to local data sources, our system uses global schema and global ontology of sources to validate and transform global XML query directly into local query formats. In addition, we use integrity constraints and validation rules for source selection and query optimization.

## III. CONSTRAINT-BASED GLOBAL SCHEMA AND MAPPINGS

In this section, global schema created by [1] is described which is used by our query distribution framework.

For creating global schema, a mapping document named as *ontology extractor changes document* is used which helps in handling conflicts between different local schema. The only constraints used in this global schema are the primary key and foreign key.

We extend this global schema creation approach by adding integrity constraints and validation rules in the global schema and mapping documents that are later used for source selection and query optimization. Three types of constraints are considered in our system which are length, format, and validation rule. Length constraint is related to the maximum length of data item defined as element in a schema. For resolving the length conflicts between different schema elements, we use maximum length of all elements being matched. Format constraint is related to the format of data in the element. We use "a" for character and "9" for a number for describing formats. In case of a data type conflict, more general type is used, for example, department_id can have number data type in one schema and character data type in another schema, so the global schema will use character data type which is a more general type. Validation rule constraints restricts the value of data in an element. For example, data for "age" element can be restricted by validation rule "lt30", which means student age must be less than 30 years at the time of enrollment. Conflicts between validation rules are handled by using union of validation rules of all involved elements.

### An Example of Constraint-Based Global Schema and Mappings Documents

An example of global schema created in [1] and extended by our system is presented here. As in this work we are not interested in creating global schema and mappings document, so details of local schema and global schema creation procedure can be seen in the example in [1] which uses two local sources described as, *Schema 1*(Student, Department, Course, Lecturer, Student_course, Lecturer_course), and *Schema 2*(Student, Course, CS_Professor, Student_Course). We extend global schema by adding constraints in it.

Based on our constraints document, we change global schema and mappings document created in [1] to add length, format and validation rule in addition to primary and foreign key constraints. An example of mappings document is shown in Listing 1. Global schema extended with constraints is shown in Listing 2 which contains elements needed for our example. As we can see from the mappings document and the global schema that global schema uses more general names and formats for schema elements to incorporate all the local schema restrictions in it.

```xml
<?xml version="1.0"?>
<OntologyExtractorChanges>
<OQL>
<Concept CDM_name="Student" ontology_name="Student">
<attribute CDM_name="student_id" ontology_name=
"registration_number" CDM_type="text" ontology_type=
"string" length="7" format="aa99999" rule="null"/>
<attribute CDM_name="fname" ontology_name="name"
CDM_type="text" ontology_type="string" length="25"
format="null" rule="null"/>
<attribute CDM_name="batch" ontology_name="batch_no"
CDM_type="text" ontology_type="text" length="4"
format="aa99" rule=""/>
<attribute CDM_name="depid" ontology_name="department_id"
CDM_type="number" ontology_type="decimal" length="null"
format="99" rule="lt12"/>
</Concept>
<Concept CDM_name="Department" ontology_name="Department">
<attribute CDM_name="dep_id" ontology_name="depid"
CDM_type="number" ontology_type="decimal" length="null"
format="99" rule="lt12"/>
<attribute CDM_name="depname" ontology_name="name"
CDM_type="text" ontology_type="string" length="4"
format="aaaa" rule="null"/>
</Concept>
</OQL>
...
</OntologyExtractorChanges>
```

Listing 1. Local source to global mappings example

Constraints added in these documents help in our query distribution framework for matching the source of data that can answer query. It can also help in optimizing the query performance by sending only the parts of a disjunctive query to sources whose constraints match with query.

```xml
<Globalmapping>
<relation name="Student">
    <PKAttribute name="registration_number"
    type="string" length="8" format="aa999999"
    rule="null"/>
    <attribute name="name" type="string"
    length="30" format="null" rule="null"/>
    <attribute name="batch_no" type="string"
    length="10" format="null" rule="null"/>
    <FKAttribute name="department_id" type="decimal"
    length="null" format="null" rule="lt20"/>
  </relation>
  <relation name="Department">
    <PKAttribute name="department_id" type="decimal"
    length="null" format="null" rule="lt20"/>
    <attribute name="name" type="string"
    length="20" format="" rule="null"/>
  </relation>
</Globalmapping>
```

Listing 2. Extended Global Schema

## IV. QUERY DISTRIBUTION FRAMEWORK

Query distribution framework is shown in Figure 1. As described earlier, our system uses an already created global schema of sources that we have extended by adding constraints.

Our query distribution framework consists of three layers; queries generation layer, local result generation layer, and final result generation layer. In the following we describe each of these layers, their components, and functionalities.

### A. Queries Generation Layer

This layer is responsible for parsing, validation, and reformulation of the global query into local queries. It uses global schema of sources which is an XML document to validate the query. A global ontology XML document is used to map local source schema with global query. This layer consist of following main components which are described below; value splitter and value attachers, scanning, parsing and validation, global query generator, and local query generators.

**Value Splitter/Attachers, and Scanning, Parsing and Validation**

Value Splitter is used to split the condition elements from the query. These condition elements are saved as objects in a list of objects as shown in Figure 1. These objects are separately translated into local source schema format when the query is reformulated using mapping document. At the end of queries generation layer, after translation of condition objects into source format, Value Attacher component attaches condition objects to the local queries.

Global XQuery is scanned, parsed and validated before reformulating it into local queries. Parsing is used to recognize and validate the query syntax [13]. Query is validated using the global schema XML document. Query elements names, i.e., table and column names must conform to names in global schema because this integrated schema and ontology is shown to user before posing the query. After validation of schema element names, constraints are validated. If any of the query element or condition value is violating the global schema element constraints, query is declared as invalid.

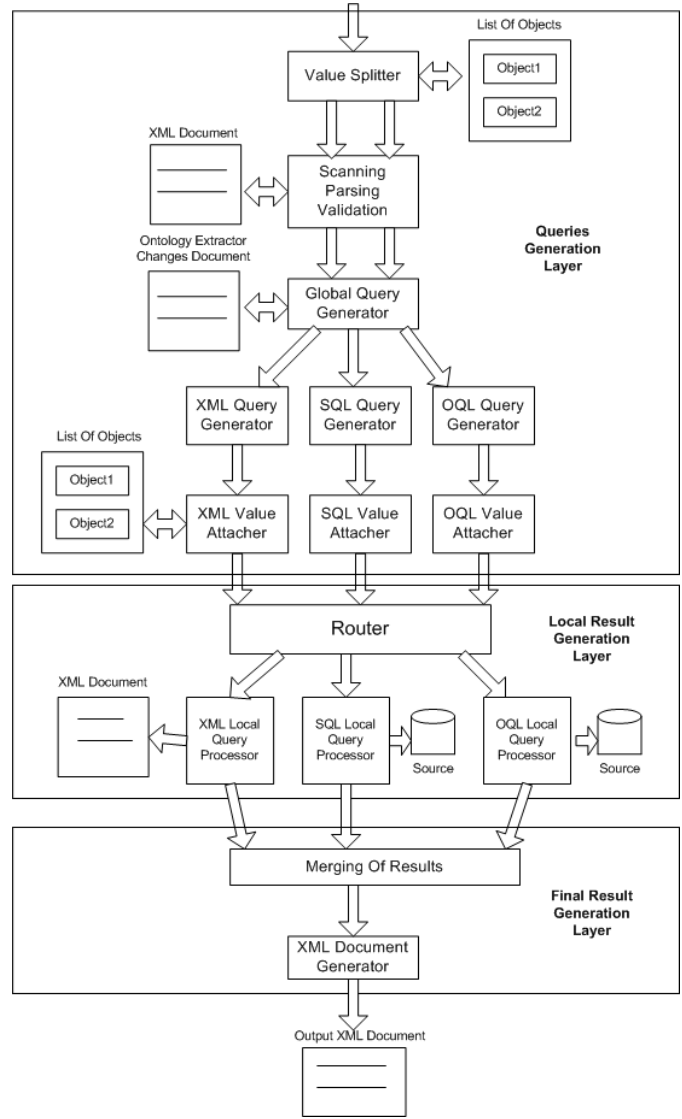

Fig. 1. Query Distribution Framework

**Global Query Generator**

Global Query Generator component performs two main functions: mappings from global to local schema, and source selection as well as query optimization. Here we describe the mappings from global to local schema. A global query posed by a user over the integrated schema needs to be reformulated according to local source schema format. Mappings between each local source schema and integrated schema is used to reformulate global query into local query format. Our system uses a mapping document for this purpose which is called *Ontology Extractor Changes Document*. This mapping document is based on ontologies of the local schemas and global schema. Mapping document provides the actual schema element names in local sources for each global element name used in query. So the query is reformulated by replacing global names of data items with local source data item names. Different methods have been used in literature for mapping from global to local query. Source ontology and broker ontology are used by [14] for this purpose. Global integrity constraints are used by [2] to reformulate the query. Mappings between local and global ontologies have been used for reformulation of query by [11]. XML DTD-based metadata dictionary is used as global ontology of sources by [9].

**Source Selection and Query Optimization**

Query is not sent to all sources for execution. It can reduce query processing performance in case of large number of sources. Sources must be selected beforehand which are capable of answering query. Source selection is performed using constraints added in the mapping documents. Global query generator component performs these tasks. Query values and conditions are checked against the local schema constraints using the mappings document. Source selection is performed when a constraint in local schema does not match with query conditions, so query is not sent to this source. Additionally, in case of a disjunctive query, if one disjunctive part in query condition is violating source schema constraints but other parts of query condition are according to source constraints then the part of query which violates source constraints is not sent to local source.

Constraints are also helpful in optimizing a query for the fragmented and replicated sources. Our system finds the replicated sources of data using constraint matching and send the query to only one of them. Fragmented sources, for example, horizontal fragmentation according to departments in our example, can be handled easily using constraints.

**Local Query Generators**

Query generators for all type of sources are used to generate local queries. Query generator takes query with local source schema names as input from the global query generator and reformulates it into local query using local query model.

Our system uses three types of sources XML, object-oriented, and relational, so we use three types of query generators using relevant query languages XML, OQL and SQL. Query generators generate query syntax in relevant languages and use local source schema names as provided by the mappings document.

*B. Local Result Generation Layer*

Local queries are sent to sources where they are executed by local query processor and local results are generated. Query distribution component/router sends queries over the network to sources. Query processor at local source understands local query, executes it on the source and generates local results. Query processors related to relational, object-oriented and XML sources are used. They takes the local queries in their local format and execute them on local source. The local results are generated from each source in their local format.

*C. Final Result Generation Layer*

Local results are retrieved in their own heterogeneous formats which may not be understandable to the user. In addition results from different sources can contain duplicate data. This layer integrates local generated results into a single XML document format. For this purpose it first converts all the results into XML format and then integrates them in a single XML document. Merging of results and XML document generation are two components in this layer.

Each of the local result source provides data in local format such as XML, relational, and object-oriented. These heterogeneous local result sets are merged and placed in a uniform format. Local results are integrated in a tabular form. Local results contain local data element names which must be changed according to the global schema as used in the user query otherwise user will not be able to understand the result of his query. Final results collected in the tabular form are converted into XML document during this phase.

User gets this document as result of his global query and is unaware of the heterogeneity, distribution and query reformulation details. Hence it is an all XML system, user poses XML query in XQuery language and gets the result as XML document.

## V. CASE STUDY

In this section, we present the working of our query distribution framework using the university student's information system as an example. We describe the working of three layers of the query distribution framework with running example. This example will also show the use of constraints added by us in the global schema and mapping document which help in source selection and query optimization. Three local sources XML, relational, and object-oriented are used in this example. Our extended global schema already presented in Section III is used to pose the query. Mapping document showing local to global schema mapping is described in Listing 3.

*A. Queries Generation Layer*

User poses an XPath query to find the registration numbers of those students who are members of batch number cs08 or members of department number 13. The actual XPath query is shown here:

*/student[batch_no="cs08" or department_id = 13] /registration_no*

Value Splitter splits the condition elements (element, sub element, element operator, element value and operator) which are saved in list of objects shown as two objects in the following table.

| | | |
|---|---|---|
| **Object1** | Element | Student |
| | Sub Element | batch_no |
| | Element Operator | = |
| | Element Value | cs08 |
| | Operator | or |
| **Object2** | Element | Student |
| | Sub Element | department_id |
| | Element Operator | = |
| | Element Value | 13 |
| | Operator | null |

Selection attributes, projection attributes, constraints on attributes as well as the concept name are separated from the query and sent for validation using global schema. Attribute values in query are matched with attribute constraints in global schema. If any violation of names or constraints is found then query is not sent for execution. After separating selection attributes, remaining query is shown here.

*/student/registration_no*

After validation of attributes and constraints from global schema XML document, global query needs to be rewritten according to source schemas. Mappings document is used for this purpose. This document contains mappings between local sources and global ontology. Mappings of three local sources, i.e, XML, relational, and object-oriented are used in our example and shown in Listing 3.

```
<?xml version="1.0"?>
<OntologyExtractorChanges>
<XML>
<Concept CDM_name="Student" ontology_name="Student">
<attribute CDM_name="id" ontology_name=
"registration_number" CDM_type="text" ontology_type=
"string" length="8" format="aaaaaaaa" rule="null"/>
<attribute CDM_name="batchno" ontology_name="batch_no"
CDM_type="text" ontology_type="text" length="4"
format="aaaa" rule=""/>
<attribute CDM_name="did" ontology_name="department_id"
CDM_type="number" ontology_type="decimal" length="null"
format="99" rule="lt20"/>
</Concept>
</XML>
<SQL>
<Concept CDM_name="Students" ontology_name="Student">
<attribute CDM_name="regno" ontology_name=
"registration_number" CDM_type="text" ontology_type=
"string" length="8" format="aa999999" rule="null"/>
<attribute CDM_name="batch" ontology_name="batch_no"
CDM_type="text" ontology_type="text" length="4"
format="aaaa" rule=""/>
<attribute CDM_name="dep_id" ontology_name="department_id"
CDM_type="number" ontology_type="decimal" length="null"
format="99" rule="lt15"/>
</Concept>
</SQL>
<OQL>
<Concept CDM_name="Student" ontology_name="Student">
<attribute CDM_name="student_id" ontology_name=
"registration_number" CDM_type="text" ontology_type=
"string" length="7" format="aa99999" rule="null"/>
<attribute CDM_name="bat_no" ontology_name="batch_no"
CDM_type="text" ontology_type="text" length="4"
format="aa99" rule=""/>
<attribute CDM_name="depid" ontology_name="department_id"
CDM_type="number" ontology_type="decimal" length="null"
format="99" rule="lt12"/>
</Concept>
</OQL>
</OntologyExtractorChanges>
```

Listing 3. An example of Mappings Document showing three local sources

Global query needs to be reformulated to local query formats so that it can be executed by local query processors. Local query generators are used to generate the queries according to local source format. Concept names/elements, selection attributes/condition elements, and projection attributes/sub elements of three local sources are taken from the mapping document as shown in the table below and sent to three local query generators.

Before creating queries from these local mappings, query conditions must be checked against constraints in local schema attributes to know whether the source is capable of answering query or any part of query. Source selection and query optimization is performed based on constraints added

| | Element | Student |
|---|---|---|
| **XML source** | Sub Element | id |
| | Condition Element | batchno |
| | Condition Element | did |
| | Element | Students |
| **SQL source** | Sub Element | regno |
| | Condition Element | batch |
| | Condition Element | dep_id |
| | Element | Student |
| **OQL source** | Sub Element | student_id |
| | Condition Element | bat_no |
| | Condition Element | depid |

in global schema and mapping document. If constraints show that a source does not have data related to query then source selection is performed and the query is not generated for this source. In addition, if a query is multi-condition query having disjunctive conditions, query is generated for only relevant part whose condition satisfies the source constraints. It is visible from query conditions in our example and mapping documents shown in Listing 3 that all three sources are capable of executing at least one part of disjunctive query.

Three query generators generate queries according to local data model. These queries do not contain query conditions because condition operators and values have been saved in objects and will be inserted in next step. Three local queries generated are shown here.

*XML query: /student/id*

*SQL query: Select regno from students*

*OQL query: Select s.student_id from s in Student*

In our example, the query is disjunctive query having two parts. It is visible from query conditions and mapping document shown in Listing 3 that the value of department_id required by the query is 13 while the condition in object-oriented source for depid is less than 12 described as "lt12". It is visible that the object-oriented source does not have the values for department 13. The query is generated for object-oriented source without the depid condition. In this way, query optimization saves execution time for part of query which the source is incapable of answering.

Value attacher component attaches the condition to the local queries using the condition operator and values from the list of objects that was saved at the start of the query processing. Thus the local query generation phase is completed and the queries are ready to be sent to sources. These local queries are shown here.

*XML query: /student[batchno="cs08" or did=13]/id*

*SQL query: Select regno from students Where batch="cs08" or dep_id=13*

*OQL query: Select s.student_id from s in Student Where s.bat_no="cs08"*

## B. Local Result Generation Layer

Local queries generated in queries generation layer are executed at sources and local results are collected from sources in this layer. Components of local result generation layer with this case study are shown in Figure 2.

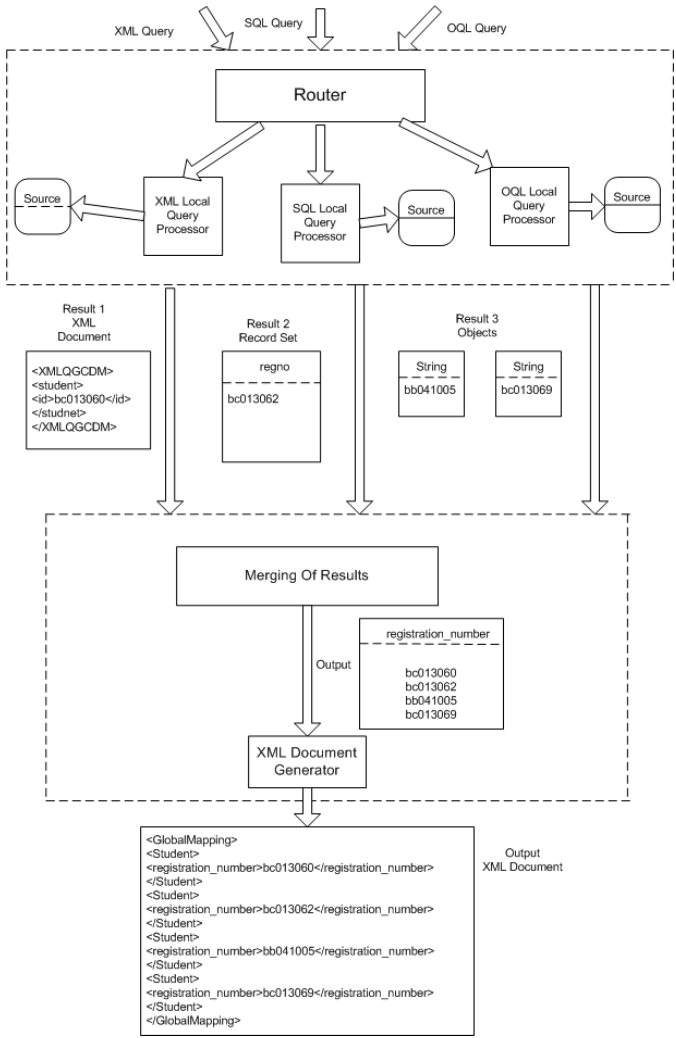

Fig. 2. Query Processing and result generation example

Local results are retrieved in format of local data sources as shown by *Result 1 XML Document*, *Result 2 Record Set*, and *Result 3 Objects* in Figure 2.

## C. Final Result Generation Layer

This layer merges heterogeneous results into XML Document format using global schema names from mapping document. Components of this layer with our case study example are shown in Figure 2. We present the results to user in XML document format. *Merging of Results* component merges local results into table format. After this merger XML document generator generates result XML document from this merged result table. This final XML result document is presented to user as the final query result. In this way user gets result in same format and using same global schema names in which he posed the query.

## VI. CONCLUSION

Querying heterogeneous data sources using all XML-based system is described in this paper. Constraints have been added in an existing global schema and mapping document. Constraints help in our query distribution framework for source selection and query optimization. The paper describes the three layers of our query distribution framework and their components.

Future work includes the evaluation of this system against benchmarks for measuring the performance of system and increasing the capability of system for handling Web-based and other types of sources.

REFERENCES

[1] Z. Huma, M. J.-U. Rehman, and N. Iftikhar, "An ontology-based framework for semi-automatic schema integration," *J. Comput. Sci. Technol.*, vol. 20, no. 6, pp. 788–796, 2005.
[2] N. Leone, G. Greco, G. Ianni, V. Lio, G. Terracina, T. Eiter, W. Faber, M. Fink, G. Gottlob, R. Rosati, D. Lembo, M. Lenzerini, M. Ruzzi, E. Kalka, B. Nowicki, and W. Staniszkis, "The infomix system for advanced integration of incomplete and inconsistent data," in *SIGMOD '05: Proceedings of the 2005 ACM SIGMOD international conference on Management of data*. New York, NY, USA: ACM, 2005, pp. 915–917.
[3] I. Kotsiopoulos, J. Keane, M. Turner, P. Layzell, and F. Zhu, "Ibhis: integration broker for heterogeneous information sources," in *Computer Software and Applications Conference, 2003. COMPSAC 2003. Proceedings. 27th Annual International*, Nov. 2003, pp. 378–384.
[4] R. Domenig and K. R. Dittrich, "A query based approach for integrating heterogeneous data sources," in *CIKM '00: Proceedings of the ninth international conference on Information and knowledge management*. New York, NY, USA: ACM, 2000, pp. 453–460.
[5] A. Goni, E. Mena, and A. Illarramendi, "Querying heterogeneous and distributed data repositories using ontologies," in *In Proceedings of the 7th European-Japanese Conference on Information Modelling and Knowledge Bases (IMKB'97*. Press, 1997.
[6] A. Halevy, A. Rajaraman, and J. Ordille, "Data integration: the teenage years," in *VLDB '06: Proceedings of the 32nd international conference on Very large data bases*. VLDB Endowment, 2006, pp. 9–16.
[7] H. H. Hoang and A. M. Tjoa, "The state of the art of ontology-based query systems: A comparison of existing approaches," in *In Proc. of ICOCI06*, 2006.
[8] A. Y. Levy, A. Rajaraman, and J. J. Ordille, "Querying heterogeneous information sources using source descriptions," in *VLDB '96: Proceedings of the 22th International Conference on Very Large Data Bases*. San Francisco, CA, USA: Morgan Kaufmann Publishers Inc., 1996, pp. 251–262.
[9] N. Arch-int, Y. Li, P. Roe, and P. Sophatsathit, "Query processing the heterogeneous information sources using ontology-based approach," in *Computers and Their Applications*, N. C. Debnath, Ed. ISCA, 2003, pp. 438–441.
[10] I. Manolescu, D. Florescu, and D. Kossmann, "Answering xml queries on heterogeneous data sources," in *VLDB '01: Proceedings of the 27th International Conference on Very Large Data Bases*. San Francisco, CA, USA: Morgan Kaufmann Publishers Inc., 2001, pp. 241–250.
[11] H. H. Hoang, A. M. Tjoa, and M. T. Nguyen, "Ontology-based virtual query system for the semanticlife digital memory project: Concepts, designs and implementation," in *4th IEEE International Conference on Computer Sciences, (RIVF 2006), HoChiMinh City, Vietnam; 12-02-2006 – 16-02-2006*. Studia Informatica Universalis, 2006, pp. 39–45.
[12] I. Manolescu, D. Kossmann, D. Florescu, D. Kossmann, F. Xhumari, and D. Olteanu, "Agora: Living with xml and relational," in *In Proceedings of International Conference on Very Large Databases (VLDB*. Morgan Kaufmann, 2000, pp. 623–626.
[13] V. Borkar, M. Carey, D. Lychagin, T. Westmann, D. Engovatov, and N. Onose, "Query processing in the aqualogic data services platform," in *VLDB '06: Proceedings of the 32nd international conference on Very large data bases*. VLDB Endowment, 2006, pp. 1037–1048.
[14] J. McHugh, S. Abiteboul, R. Goldman, D. Quass, and J. Widom, "Lore: A database management system for semistructured data," *SIGMOD Record*, vol. 26, pp. 54–66, 1997.